\documentclass{hep99}
\usepackage{graphicx}
\begin{document}

\title{Tests of QCD using differences between
gluon and quark jets$^{\dag}$}
\author{\vspace*{-.5cm}
J. William Gary}
\address{Department of Physics, University of California, 
  Riverside, CA, 92521, USA\\[3pt]
E-mail: {\tt william.gary@ucr.edu}}
\abstract{I present recent results from LEP
which utilize differences
between gluon and quark jets to make quantitative tests of QCD.
The principal topic is a determination of the ratio of QCD
color factors, C$_{\mathrm{A}}$/C$_{\mathrm{F}}$,
using either the multiplicity or fragmentation 
functions of the jets.  
In addition,
I discuss a recent measurement of the rate of
$\eta$ mesons in gluon jets compared to quark jets.
\\[6pt]}

\maketitle


\section{Introduction}
The physics of differences between gluon and quark jets
has enjoyed a golden age at LEP
due to the large data samples
and good detector capabilities.
Starting with samples of symmetric three jet ``Y'' events,
in which the two lowest energy jets both form an angle of
about $150^\circ$ with respect to the 
highest energy jet~\cite{bib-opal91},
the LEP experiments established the basic
phenomenology of this field:
gluon jets are broader, 
have a softer fragmentation function
and a larger mean multiplicity
than quark jets~\cite{bib-qgdifflep}.
These discoveries, made in 1991-1996,
culminated experimental efforts that had unsuccessfully
attempted to establish these differences since the late 1970s.
The field has now evolved beyond exposition of
gluon-quark jet differences
to quantitative tests of QCD.
Amongst these are tests of analytic calculations
of higher moments of the multiplicity distributions
of gluon and quark jets~\cite{bib-opal97,bib-lupia}
and determinations of the ratio of QCD color factors, 
C$_{\mathrm{A}}$/C$_{\mathrm{F}}$.
In the following I focus on the latter topic
since these results have appeared just this past year.
There are two studies I will discuss,
presented in OPAL publication CERN-EP/99-028
(in press in Eur. Phys. J~{\bf{C}})
and in DELPHI paper DELPHI 99-127 Conf.~314.
Also note a recent DELPHI publication~\cite{bib-delqg}
on a related topic.
I will also discuss a new result presented in
OPAL Physics Note PN407
on the production rate of $\eta$ mesons in gluon
and quark jets.

\section{Gluon and quark jets from a point source}
\label{sec-gincl}

In QCD calculations,
quark and gluon jets are produced as virtual
q$\overline{\mathrm{q}}$ and gg pairs, respectively,
from a color singlet point source.
The jet properties are defined by an inclusive sum over 
event hemispheres.
The hemispheres are defined by the plane perpendicular 
to the principal event axis.
For jets defined in this manner,
referred to as ``unbiased,''
there is no jet finding algorithm and no ambiguity
about which particles to associate with gluon or
quark jet production.
Experimental access to high energy unbiased quark jets
is easy since hadronic events in e$^+$e$^-$ annihilations
result from
q$\overline{\mathrm{q}}$ production from
a color singlet source.
In contrast,
gg production from a color singlet point source
is a process which has been practically unobserved in nature.
One channel where the experimental selection of gluon jets
matches the theoretical criteria is e$^+$e$^-$
hadronic annihilation events
in which the quark jets q and 
$\overline{\mathrm{q}}$ from the electroweak 
Z$^0/\gamma$ decay are approximately colinear:
the gluon jet hemisphere against which the 
q and $\overline{\mathrm{q}}$ recoil is produced under the same
conditions as gluon jets in gg
events~\cite{bib-valery88,bib-gary94}.
OPAL selected events 
of the type e$^+$e$^-$$\rightarrow\,$q$_{\mathrm{tag}}
\overline{\mathrm{q}}_{\mathrm{tag}}$g$_{\mathrm{incl.}}$,
in which g$_{\mathrm{incl.}}$ refers to a gluon jet hemisphere
recoiling against two tagged quark jets q$_{\mathrm{tag}}$
and $\overline{\mathrm{q}}_{\mathrm{tag}}$ in the
opposite hemisphere.
Monte Carlo study shows that these
gluon jet hemispheres have almost
identical properties to unbiased gluon jets
and can be selected with virtually no dependence
on a jet finding algorithm,
a unique feature of this method.
The OPAL results are
obtained for jet energies of 40~GeV.

\fntext{\dag}{Talk given at the International Europhysics Conference,
High Energy Physics 99,
Tampere, Finland, 15-21 July 1999.}

\begin{figure}[h]
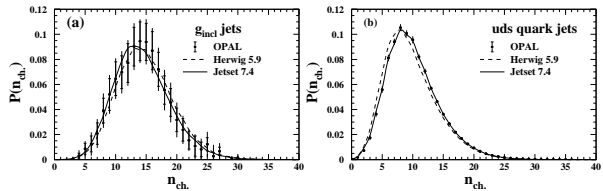

\begin{center}
\begin{tabular}{cc}
  \hspace*{-.4cm}\includegraphics[scale=.25]{corrected_data_parta.epsi} &
  \hspace*{-.4cm}\includegraphics[scale=.25]{corrected_data_partb.epsi} \\
 \end{tabular}
\end{center}
\vspace*{-.5cm}
\caption{Charged particle multiplicity of unbiased (a)~gluon 
and (b)~uds flavored quark jets,
defined by inclusive sums over event 
hemispheres~\cite{bib-opal97}.
}
\label{fig-gqmult}
\vspace*{-.5cm}
\end{figure}

The charged particle multiplicity distributions 
of the unbiased gluon and quark jets
are shown in Fig.~\ref{fig-gqmult}a and b,
respectively~\cite{bib-opal97}.
The measured ratio of the mean multiplicity between gluon
and quark jets is
$1.51\pm 0.02\,(\mathrm{stat.})\pm 0.04(\mathrm{syst.})$.
This is in excellent agreement with recent QCD calculations
of this quantity~\cite{bib-lupia-ochs}.

\begin{figure}[h]
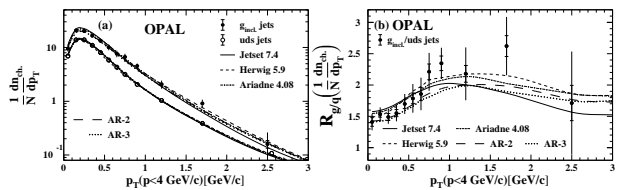

\begin{center}
 \begin{tabular}{cc}
   \hspace*{-.4cm}\includegraphics[scale=.25]{pt_ptotlt4gev_parta.epsi} &
   \hspace*{-.4cm}\includegraphics[scale=.25]{pt_ptotlt4gev_partb.epsi} \\
 \end{tabular}
\end{center}
\vspace*{-.5cm}
\caption{(a) $p_{T}$ distributions of 
soft charged particles
in unbiased gluon and uds flavored quark jets,
(b)~Ratio of the gluon to quark jet measurements from~(a).
}
\label{fig-gqpt}
\vspace*{-.3cm}
\end{figure}

Multiplicity in full phase space is sensitive to
higher order corrections and the effects of energy conservation.
Multiplicity in {\it limited} phase space,
such as the multiplicity of soft particles,
is much less sensitive to these effects and better
satisfies the asymptotic condition of QCD.
Thus one can potentially observe the
\underline{full color factor difference}
C$_{\mathrm{A}}$/C$_{\mathrm{F}}$=2.25
in the ratio of soft hadron multiplicities,
limited only by possible finite energy and
hadronization effects.
Indeed it has been predicted that the ratio of
soft particles at large transverse momentum
$p_{T}$ between gluon and quark jets should
approximately equal 2.25 even at the finite
energies of LEP~\cite{bib-khoze98}.
Fig.~\ref{fig-gqpt}a shows the $p_{T}$ spectrum
of soft particles in the unbiased gluon and quark jets.
$p_{T}$ is defined relative to the jet axes
obtained by summing the particle 3-momenta in the gluon
and quark jet hemispheres.
Soft particles are defined by momenta $p$$<$4.0~GeV/$c$.
Fig.~\ref{fig-gqpt}b shows the ratio of the gluon
to quark jet results.
For $p_{T}$ larger than about 0.7~GeV/$c$,
the ratio of the
gluon to quark jet multiplicity is indeed about 2.25.
Integrating the results of Fig.~\ref{fig-gqpt}b between
0.8 and~3.0~GeV/$c$ yields
\begin{equation}
 r = 2.29\pm0.09\,(\mathrm{stat.})\pm0.15(\mathrm{syst.})
 \label{eq-rch}
\end{equation}
as the multiplicity ratio of soft particles at large $p_T$.
This result is summarized in the top row of Table~1.

\begin{table}
\label{tab-rpt}
\begin{center}
\caption{Results for the multiplicity ratio between
gluon and quark jets for soft particles with large
transverse momentum with respect to the jet axes.
}
\begin{tabular}{ll}
\br
OPAL data & $2.29\pm0.17$\\ 
\mr
HERWIG hadrons, E$_{\mathrm{c.m.}}$=91~GeV & 2.16 \\ 
HERWIG partons, E$_{\mathrm{c.m.}}$=91~GeV & 2.09 \\
HERWIG hadrons, E$_{\mathrm{c.m.}}$=10~TeV & 2.24 \\ 
HERWIG partons, E$_{\mathrm{c.m.}}$=10~TeV & 2.25 \\ 
JETSET partons, E$_{\mathrm{c.m.}}$=91~GeV, &  \\
   C$_{\mathrm{A}}$=C$_{\mathrm{F}}$=4/3 & 1.00 \\ 
\br
\end{tabular}
\end{center}
\vspace*{-.7cm}
\end{table}

One can ask if this result truly constitutes a
measurement of C$_{\mathrm{A}}$/C$_{\mathrm{F}}$.
This can be demonstrated using QCD Monte Carlo programs
since these programs incorporate QCD by definition
and should yield C$_{\mathrm{A}}$/C$_{\mathrm{F}}$=2.25
under the appropriate circumstances.
The HERWIG Monte Carlo contains
next-to-next-to-leading-order QCD plus energy-momentum
conservation at each branching and exhibits the correct
asymptotic behavior and thus is appropriate for this test.
If the experimental analysis is applied to HERWIG events
at the parton level at asymptotically large energies
(e.g.~10~TeV),
the result should equal 2.25 if the analysis
truly measures C$_{\mathrm{A}}$/C$_{\mathrm{F}}$.
Similarly one can use parton level Monte Carlo events with
C$_{\mathrm{A}}$=C$_{\mathrm{F}}$=4/3 to verify that
the experimental analysis yields unity,
again a necessary condition if the analysis
truly measures C$_{\mathrm{A}}$/C$_{\mathrm{F}}$.
For the latter test, the JETSET Monte Carlo is used
since HERWIG does not allow C$_{\mathrm{A}}$=C$_{\mathrm{F}}$.
The Monte Carlo results for the experimental
variable $(r_{\,\mathrm{ch.}})^{p<4\,{\mathrm{GeV}}/c}
           _{0.8<p_T<3\,{\mathrm{GeV}}/c}$ are
given in the bottom portion of Table~1.
The results are 2.25 for HERWIG at the parton level
with E$_{\mathrm{c.m.}}$=10~TeV and
1.00 for JETSET at the parton level
with C$_{\mathrm{A}}$=C$_{\mathrm{F}}$=4/3.
This demonstrates that the result (\ref{eq-rch}) is indeed
a measurement of C$_{\mathrm{A}}$/C$_{\mathrm{F}}$:
it is in fact one of the most accurate measurements
of this ratio yet performed.
Note that the Herwig results for E$_{\mathrm{c.m.}}$=91~GeV
in Table~1 imply a hadronization correction of about 0.97
to the measurement~(\ref{eq-rch}).

Beyond the measurement of C$_{\mathrm{A}}$/C$_{\mathrm{F}}$
discussed above,
OPAL uses their g$_{\mathrm{incl.}}$ jet sample to exclude
the AR-2 and AR-3 models of color reconnection implemented in the
ARIADNE QCD Monte Carlo program with a significance
of about five standard deviations.
This is currently the most stringent limit on any realistic
model of color reconnection.

\section{C$_{\mathrm{A}}$/C$_{\mathrm{F}}$ from the scale
dependence of gluon and quark jet fragmentation functions}

The jets discussed in the previous section have
a fixed energy of about 40~GeV.
It is of interest to obtain information about
the internal properties of gluon jets at other scales,
using three jet q$\overline{\mathrm{q}}$g
events selected in a standard way with a jet finding algorithm.
Higher order terms associated with the phenomenon
of coherence suggest that the characteristics of a jet
depend on a transverse momentum-like scale,
$\kappa$=E$_{\mathrm{jet}}
\sin(\theta_{\mathrm{min.}}/2)$~\cite{bib-kappa},
where E$_{\mathrm{jet}}$ is the energy of the jet
and $\theta_{\mathrm{min.}}$ is the smaller of the
angles with respect to the other two jets.
It has previously been demonstrated that $\kappa$ 
can be an appropriate variable for comparing jets in 
different three jet topologies~\cite{bib-alephtopology}.

DELPHI has examined the scale dependence of gluon and
quark jet fragmentation functions in three jet events.
Three jet events are selected using either the $k_\perp$
or Cambridge jet finders.
The impact parameters of charged tracks are used to
identify gluon jets within those events.
The fragmentation functions of the two lower energy jets,
one of which is a quark jet and the other a gluon jet
with high probability,
are measured as a function of~$\kappa$.
The results for quark jets are shown in
Fig.~\ref{fig-ffquark}.
\begin{figure}[ht]
\begin{center}
  \includegraphics[scale=.23]{delphi_quarkff.epsi}
\end{center}
\vspace*{-.5cm}
\caption{Quark jet fragmentation function vs. scale.
}
\label{fig-ffquark}
\vspace*{-.3cm}
\end{figure}
The DELPHI results for quark jets are in good
agreement with the unbiased (hemisphere) results
from the TASSO and TPC/2$\gamma$ experiments.
This agreement supports the appropriateness of $\kappa$ as the
scale for this analysis.
Fig.~\ref{fig-ffgluon} shows the corresponding
results for gluon jets.
\begin{figure}[h]
\begin{center}
  \includegraphics[scale=.23]{delphi_gluonff.epsi}
\end{center}
\vspace*{-.5cm}
\caption{Gluon jet fragmentation function vs. scale.
}
\label{fig-ffgluon}
\vspace*{-.3cm}
\end{figure}
The DELPHI results for gluon jets
extrapolate well to the OPAL results for the fragmentation
function of unbiased gluon jets at $\kappa$$\approx$40~GeV
(CERN-EP/99-028).
This again supports the appropriateness of
$\kappa$ as the scale.

The scale evolution of fragmentation functions in QCD
is described by the DGLAP equations.
DELPHI parametrizes the fragmentation functions of 
the gluon and quark jets at a fixed scale,
chosen to be $\kappa$=5.5~GeV.
A simultaneous fit is made
to the DELPHI data in Figs.~\ref{fig-ffquark}
and~\ref{fig-ffgluon},
using first order DGLAP evolution,
to fix the variables of
this parametrization and to determine the color
factor C$_{\mathrm{A}}$ and the effective
QCD scale parameter $\Lambda$.
This strategy is very similar to that employed 
in deep inelastic scattering to determine
the strong coupling strength, $\alpha_S$,
from the scale evolution of structure functions.
The results for C$_{\mathrm{A}}$ and $\Lambda$
are $2.97\pm0.12$~(stat.) 
and $0.40\pm 0.11\,$GeV~(stat.),
respectively.
Dividing the result for C$_{\mathrm{A}}$ by
C$_{\mathrm{F}}$=4/3 yields
C$_{\mathrm{A}}$/C$_{\mathrm{F}}$=$2.23\pm0.09\,
{\mathrm{(stat.)}}\pm0.23\,{\mathrm{(syst.)}}^{\dag}$
\fntext{\dag}{The systematic uncertainty for
this result is based on the full difference
between the standard measurement and the measurements with
a systematic change in the analysis,
and not half the difference
as in DELPHI 99-127 Conf.~314,
to make it comparable to other results
for C$_{\mathrm{A}}$/C$_{\mathrm{F}}$ such as the one
presented in section~\ref{sec-gincl}.}
where the systematic uncertainty is mostly due to
the difference between using the
$k_\perp$ or Cambridge jet finders.
This result is in good agreement with the QCD value of~2.25.

\section{$\eta$ meson rates in gluon and quark jets}

QCD predicts that the ratio of the mean multiplicity in
gluon to quark jets, $r_h$,
is the same for all hadron species~$h$.
$r_h$ could differ for different types of hadrons,
however, for at least two reasons:
(1)~because of the decay properties of hadrons,
e.g.~B hadrons yield many kaons, making 
$r_{h}$ smaller for kaons
than the corresponding result for other particles
since e$^+$e$^-$ jets with B hadrons are almost 
always quark jets,
or (2)~dynamical differences between the hadronization
properties of gluons and quarks.
Examples of dynamical differences
occur in the Lund hadronization model,
which predicts that $r_h$ is larger for baryons than for mesons,
and in the gluon octet string model~\cite{bib-octetg},
which predicts $r_h$ to be larger for isoscalar
mesons than for non-isoscalar ones.

To test for the presence of a dynamical enhancement 
of isoscalar meson production in gluon jets,
L3~\cite{bib-l3eta} measured the $\eta$ meson
rate in the lowest energy jet (``jet 3'')
of three jet events and compared the result
to the corresponding Monte Carlo prediction
with no isoscalar enhancement mechanism.
They found the $\eta$ rate in jet 3 to be
$30\pm10$\% larger in the data than in the MC,
which was interpreted as evidence for the
dynamical enhancement of isoscalars in gluon jets.

OPAL has performed a related study.
Three jet events are reconstructed using either
the $k_\perp$, Luclus or cone algorithms.
The charged particle,
$\pi^0$ and $\eta$ multiplicities of the jets are
compared as a function of the $\kappa$ scale
discussed in the previous section.
The measurements are unfolded to correspond to
pure quark and gluon jets in the regions where
the $\kappa$ scales of jets overlap.

\begin{figure}[h]
\begin{center}
\begin{tabular}{cc}
 \includegraphics[scale=.26]{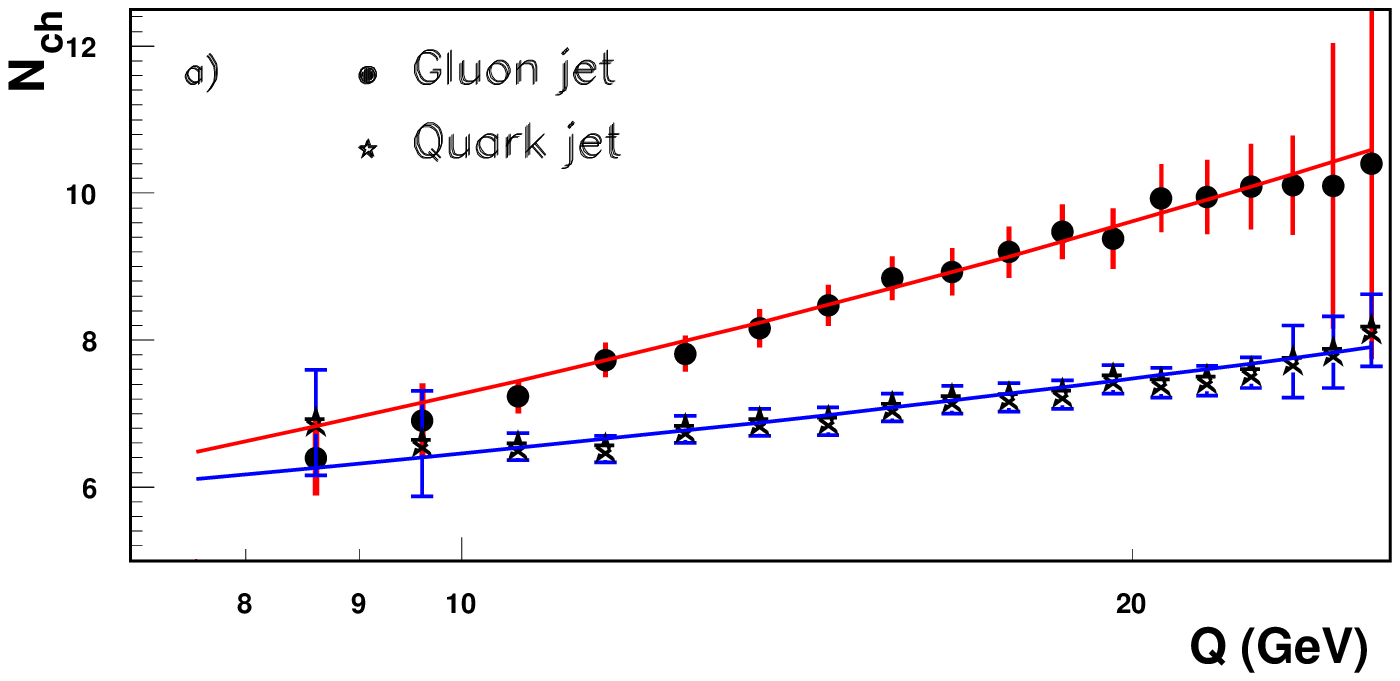} & \\[-.45cm]
 \includegraphics[scale=.26]{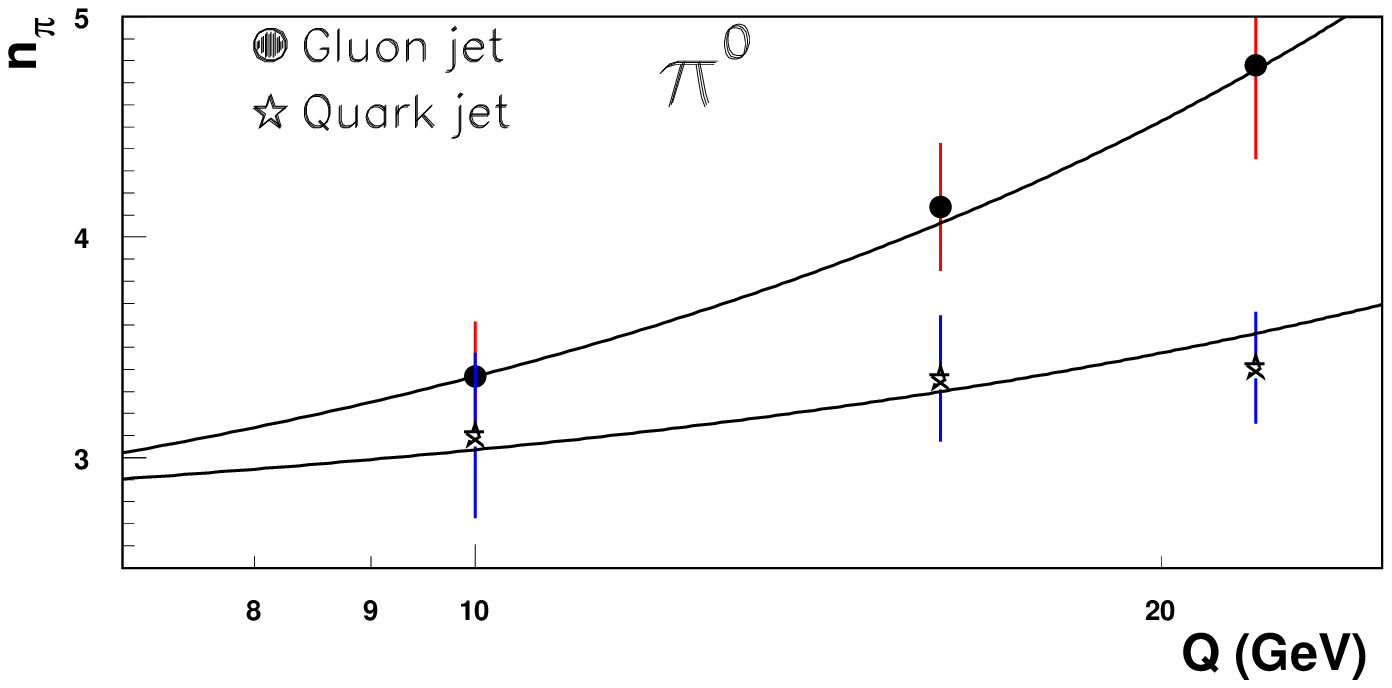} &
 \hspace*{-.2cm}\includegraphics[scale=.26]{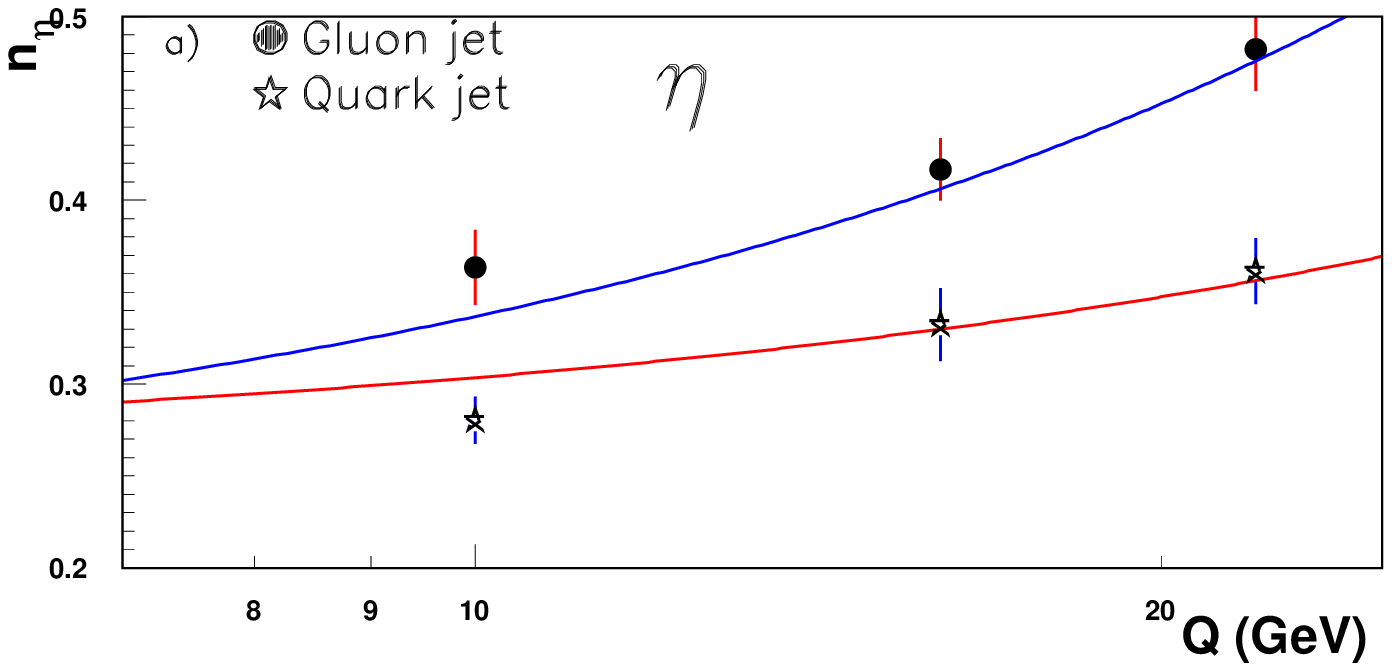} \\
 \end{tabular}
\end{center}
\vspace*{-.5cm}
\caption{(a) charged particle, (b)~$\pi^0$,
and (c)~$\eta$ meson rates in gluon and quark jets
vs. scale Q=$\kappa$,
using the Luclus jet finder.
}
\label{fig-eta}
\vspace*{-.3cm}
\end{figure}

The results for all charged particles
are shown in Fig.~\ref{fig-eta}a 
(these are very similar to recently published
data from DELPHI~\cite{bib-delqg}).
The curves in that figure show
a polynomial parametrization of the measurements.
The corresponding results for $\pi^0$ and $\eta$ mesons
are shown in Fig.~\ref{fig-eta}b and~c.
To test for dynamical differences between the production rates
of $\pi^0$ and $\eta$ mesons in gluon and quark jets
compared to charged particles,
the parametrizations from Fig.~\ref{fig-eta}a are scaled
using the same factor for both gluon and quark jets to
obtain the curves in Fig.~\ref{fig-eta}b,
and similarly for Fig.~\ref{fig-eta}c.
The scaled parametrizations
describe the $\pi^0$ and $\eta$ measurements accurately
with the exception of a slight disagreement in the lowest
bin of the $\eta$ distribution (Fig.~\ref{fig-eta}c) which is
not statistically significant.
Thus the ratios $r_h$ for $\pi^0$ and $\eta$ mesons
agree with $r_h$ for charged particles
in all bins of $\kappa$ to within the uncertainties.
Therefore OPAL does not obtain evidence for a dynamical
enhancement of $\eta$ mesons in gluon jets,
in contrast to L3.
ALEPH~\cite{bib-alepheta}
recently reported results for the $\eta$ meson rate
in jet 3 of three jet events,
similar to the L3 study.
ALEPH observes the measured result to be well
described by the Monte Carlo
without a mechanism for isoscalar meson enhancement,
in contrast to L3.
Thus --~similar to OPAL~--
ALEPH does not obtain evidence for
isoscalar meson enhancement in gluon jets.

\section{Summary}

After 20 years of experimental effort,
the field of differences between gluon and quark jets
has advanced to the level of
providing precise, quantitative tests of QCD.
The gluon to quark jet multiplicity ratio in full phase
space is 
$1.51\pm 0.02\,(\mathrm{stat.})\pm 0.04(\mathrm{syst.})$,
in agreement with QCD calculations which incorporate
energy conservation and the correct phase space limits
for soft gluon radiation~\cite{bib-lupia-ochs}.
For soft particles with large transverse momentum
with respect to the jet axes,
the corresponding result is
$2.29\pm0.09\,(\mathrm{stat.})\pm0.15(\mathrm{syst.})$,
providing one of the most accurate
measurements of the ratio of QCD color factors,
C$_{\mathrm{A}}$/C$_{\mathrm{F}}$.
C$_{\mathrm{A}}$/C$_{\mathrm{F}}$ has also been measured 
using the difference in the scale evolution of 
gluon and quark jet fragmentation functions 
to be
$2.23\pm0.09\,{\mathrm{(stat.)}}\pm0.23\,{\mathrm{(syst.)}}$.
Last,
non-perturbative aspects of gluon and quark jet differences
have been probed in
two recent studies which find no evidence for
an enhancement of $\eta$ mesons in gluon jets
compared to quark jets,
beyond that observed for charged particles.
These latter results are in contrast to the conclusions
of an earlier L3 study~\cite{bib-l3eta}.

\end{document}